\documentclass[conference,compsoc]{IEEEtran}
\ifCLASSOPTIONcompsoc
  \usepackage[nocompress]{cite}
\else
  \usepackage{cite}
\fi
\ifCLASSINFOpdf
\else
\fi
\usepackage{graphicx}
\usepackage{algorithm}
\usepackage{algpseudocode}
\usepackage{amsmath,amsfonts,amssymb,amsthm}
\begin{document}
\title{Unsupervised Learning Based Robust Multivariate Intrusion Detection System for Cyber-Physical Systems using Low Rank Matrix}

\author{\IEEEauthorblockN{Aneet K. Dutta}
\IEEEauthorblockA{C3i Center, Computer Science \& \\Engineering Department\\
Indian Institute of Technology Kanpur\\
India\\
aneet@cse.iitk.ac.in}
\and
\IEEEauthorblockN{Bhaskar Mukhoty}
\IEEEauthorblockA{Computer Science \& \\Engineering Department\\
Indian Institute of Technology Kanpur\\
India\\
bhaskarm@cse.iitk.ac.in}
\and
\IEEEauthorblockN{Sandeep K. Shukla }
\IEEEauthorblockA{C3i Center, Computer Science \& \\Engineering Department\\
Indian Institute of Technology Kanpur\\
India\\
sandeeps@cse.iitk.ac.in}
}


% compsoc conferences compared to traditional conferences), use this
%Telephone: (800) 555--1212, Fax: (888) 555--1212}


\maketitle

\begin{abstract}

Regular and uninterrupted operation of critical infrastructures such as power, transport, communication etc. are essential for proper functioning of a country. Cyber-attacks causing disruption in critical infrastructure service in the past, are considered as a significant threat. With the advancement in technology and the progress of the critical infrastructures towards IP based communication, cyber-physical systems are lucrative targets of the attackers. In this paper, we propose a robust multivariate intrusion detection system called RAD for detecting attacks in the cyber-physical systems in $O(d)$ space and time complexity, where $d$ is the number parameters in the system state vector. The proposed Intrusion Detection System(IDS) is developed in an unsupervised learning setting without using labelled data denoting attacks. It allows a fraction of the training data to be corrupted by outliers or under attack, by subscribing to robust training procedure. The proposed IDS outperforms existing anomaly detection techniques in several real-world datasets and attack scenarios.

\end{abstract}


% \fi
\IEEEpeerreviewmaketitle
\section{Introduction}

The entire economy of a nation is dependent upon the proper functioning of the critical infrastructure. The critical infrastructures include health care, manufacturing, water treatment, transportation, and power system. Even a short disruption in the service of critical infrastructures can affect public life severely. With the advancement network technology in the Supervisory Control \& Data Acquisition (SCADA)\cite{c2} systems, an essential component of any critical infrastructure, IP based communication has replaced serial communication, exposing the system to the attackers over the IP networks. With IP convergence networks got further integrated with the IT network, inviting adversaries outside operational technology (OT) networks. There is an immediate need to detect the attacks in these systems and generate an alert in real-time. In order to generate real time alerts, the IDS is constrained by low time complexity during the test phase.

The operational technology of a cyber-physical system ensures proper functioning of the system without considering the external factors like security. As a result many security vulnerabilities exist in the cyber-physical systems. For example in the industrial communication protocol like MODBUS, the packets are not encrypted and do not have any authorization mechanism to validate the sender, thus making the protocol vulnerable to Man In The Middle (MITM) attack or false data injection. The vulnerabilities like remote code execution(RCE), improper credential management, stack \& buffer overflow, and memory corruption make the SCADA software and Human Machine Interfaces(HMI) devices easy target for the attackers. Since, changing all the equipment in the existing infrastructure with updated and secured protocols are time-consuming and expensive, we developed a solution that can retrofit with the existing systems to protect the system with the detection of attacks in real-time.

There are two kind of approaches in designing an IDS, model-driven approach and data-driven approach. Model-driven based IDS consists of rules that continuously checks system dynamics are not violated. The IDS used in Industrial Control System(ICS) are generally model driven as the system can be modeled using laws of physics. Often, these systems are large and complex, consisting of many physical parameters that are measured by sensors and controlled by actuators. Manually identifying rules to model the behavior of these complex systems is often not feasible. The data-driven approach helps us in capturing ICS behavior and understanding the underlying system dynamics. In real-time, when the system functioning deviates from the modeled behavior beyond a predetermined threshold, it is considered an anomaly. 

With advancements in the field of machine learning and deep learning, many data-driven based IDS were introduced. As, such intrusion detection methods are dependent on the data, there are some challenges and concerns. These are:
\begin{itemize}
    \item The unavailability of properly labeled data classifying the cyber-physical system behavior.
    \item Most machine learning and deep learning-based IDS are not outlier and adversary resistant. Since training data might be corrupted, robustness of these detection mechanisms are not guaranteed.
    \item The use of highly sophisticated machine learning or deep learning models may not be feasible in a resource-constrained deployment environment.
\end{itemize}

In this paper, our work focuses on devising a robust anomaly detection algorithm that will learn the stable structure of the cyber-physical system and detect anomaly when the system behavior deviates from the modeled behavior learned by the algorithm. The stable behaviour of the cyber-physical system is captured in a low dimensional subspace, by applying the Robust Principal Component Analysis (RPCA) technique. The vector comprising of readings from sensors and actuators associated with the system, is projected in the low dimensional subspace and their median is computed. At the test phase, vectors are projected in the low-dimensional subspace, and based on the distance of the test vector from the median anomaly is flagged. The main contributions of this paper are:
 
\begin{itemize}
     \item We propose Robust Anomaly Detection (RAD), a multivariate model to detect anomaly in the behavior of a cyber-physical system.
     \item The proposed intrusion detection model is robust to corruption in training data (e.g poisoning attack or Gaussian noise) and it is outlier resistant. It does not require labeled training data.
     \item The IDS developed requires very limited memory in deployment and it is time efficient, making it suitable for real-time anomaly detection.
     \item The proposed method retrofits to the existing infrastructure of an ICS.
     \item We demonstrate effectiveness of RAD using several real world datasets.
\end{itemize}

\begin{figure*}[h!]
\includegraphics[width=\textwidth,height=1.5cm]{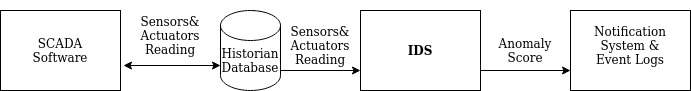}
\caption{Architecture of the integrated SCADA software \& IDS within SCADA host}
\label{fig:architecture}
\end{figure*}

\section{Background and Related Work}
With the rapid progress towards industry 4.0 and IP-convergence the risk of cyber-attacks in the cyber-physical systems are increasing every day. Since chances of state-sponsored attacks on the cyber-physical systems are high, proper detection and prevention mechanism needs to be in place to protect these infrastructures. There has been a significant amount of research going on in this area, researchers from different backgrounds have proposed intrusion detection mechanisms for the cyber-physical systems. Goldenberg et al.\cite{c2} proposes a deterministic finite automata (DFA) based intrusion detection system, where network traffic of the cyber-physical system is modeled by the DFA. However, construction of minimal-sized DFA, that reaches a accepting state when the system is under normal behavior and reaches a non-desired or dead state when the system behaves maliciously, is NP-hard. DFA based IDS are highly sensitive and have a high rate of false-positives. Cheng et al.\cite{c11} proposed an finite-state automata (FSA) based intrusion detection mechanisms similar to \cite{c2}, but it is designed to model the control flow of the program, e.g. the system calls, API calls, and memory addresses referred in the cyber-physical system. The IDS proposed by Cheng et al.\cite{c11} has significant overhead and time constraints because of continuous monitoring of the software system calls. It has a limited ability to detect attacks as it only detects when the attacker changes the execution control flow in the cyber-physical system. Adepu et al.\cite{c5} proposed an invariant based technique for generating alarms, which is activated when the system violates the invariant derived from different system variables. To generate the correct invariants, the intrusion detection developer must have detailed knowledge about the cyber-physical system dynamics.

Bernabeu et al.\cite{c1} developed an intrusion detection mechanism based on a decision tree algorithm for the power grid. They demonstrate an interesting fact about how the critical points in the power-grid changes during different environmental scenarios and the necessity of different intrusion detection models in different seasons. Hence, the supervised learning based intrusion detection mechanisms need to be updated with properly labeled data of normal and attack classes, at different environmental conditions. However, the availability of adequately labeled data of cyber-physical systems at different conditions is a considerable challenge.

Since the availability of adequately labeled data of cyber-physical system behavior is a challenge, various anomaly-based intrusion detection systems are proposed. The idea behind any anomaly-based intrusion detection mechanism is to model normal behaviour of the cyber-physical system in training time and measure the deviation from the normal behavior during the run-time. If the measure of the deviation crosses the pre-determined threshold, the anomaly is flagged. Time-series prediction using recurrent neural network (RNN) and long short term memory(LSTM) models were used for anomaly-based intrusion detection\cite{c7,c8,c14,c15}. In such cases a time-series forecasting model for each system variable is trained using the normal system behaviour. At each time step, the model predicts the next reading of the system variable. If the difference between the actual reading and the predicted reading is more than a predetermined threshold, then anomaly is flagged. 

An underlying assumption behind developing these anomaly-based IDS is that the data used for training the models is outlier free. This may not be the case in a practical setting, due to malfunctioning of different sensors, resulting corruption or missing data in the training dataset. The presence of outliers in the training set may profoundly affect performance and accuracy of the IDS. Aoudi et al. \cite{c10} proposed PASAD, an anomaly-based IDS in which for each variable, the time-series data is converted to a Hankel matrix. A low-dimensional subspace is then identified using SVD, that captures the stable structure of the cyber-physical system dynamics. Since SVD is sensitive to the outlier, the method is not robust to the outliers present in the training data. Moreover, the distance of a vector in the projected space is computed with respect to the pre-computed mean. Since, mean can be corrupted by a single outlier, the method is not robust in the training time. It is also models each sensors separately, thus multiple models need to be trained and deployed for detecting the attacks in the overall infrastructure, rendering it in-feasible in a resource constrained environment.

The ideas in \cite{c6,c13} are similar to that of in \cite{c7},\cite{c8},\cite{c14},\cite{c15} but uses a multivariate setting. The multivariate IDS solves the problem of deploying separate models for each sensor, thus can be deployed in a real-time resource-constrained environment. However, the assumptions of outlier-free training data and entire training data belonging to a normal class, persists. Therefore, these IDS can not guarantee robustness essential for a practical scenario.

\section{System Architecture}

In our work, the communication link between the PLC and SCADA is considered as the threat vector. Due to vulnerable industrial communication protocols like MODBUS, command injection, false data injection, replay, and MITM attacks are possible. It enables the attacker to change the dynamics of the system. Also, some network-based attacks on actuation signals will change the dynamics, and hence even if the sensor data is correct, but the attack now reflects in the changing dynamics of the cyber-physical system.

The attack scenarios are generated by manipulating the sensor and actuator values in the data packets communicated between the PLC and SCADA (false data injection attack). Any number of sensors, actuators or their combinations can be modified, representing attack scenarios or undesired state for the cyber-physical system. The SCADA host collects complete data of the infrastructure and maintains the overall state of the system. IDS is deployed within SCADA reads the measurement of the sensors and actuators from the database to determine whether current behavior of the cyber-physical system is normal or anomalous.

Figure \ref{fig:architecture} describes the data flow between different components within the SCADA host. All the sensors and actuators readings, along with the timestamp, are logged into the historian server. The SCADA software reads the data from the historian server to monitor the current system state. Proposed IDS deployed in the SCADA host, reads the data from the historian server in real-time, calculates the anomaly score, and sends the necessary information to the notification system, and stores it in the event log database.

The proposed IDS developed by considering the system behavioral characteristics will help to defend the infrastructure against the attackers, trying to change the state of the system to an undesirable state by manipulating the system dynamics.

\section{Proposed Methodology}
Intrusion detection in a cyber-physical system is a general anomaly detection problem. Deep learning and machine learning techniques are widely used to for anomaly detection. However, specific security issues exist in these techniques, which may enable the attacker to create carefully crafted inputs, such that the intrusion detection mechanisms fails to classify it as an anomaly. Therefore, robustness of the intrusion detection mechanisms is essential to protect the model against outliers and poisoning of the training data.

The stable behavior of a cyber-physical system can be captured by projecting the system state vectors into a low-dimensional subspace and localizing it there. Since the system state vectors are observed at multiple time instance, they in general have redundancy. Projecting them to low dimensional subspace using a clever approach helps to get rid the possible corruptions present in the data. The anomalous behavior of the system can then be detected when the system departures from the modeled behavior.

\subsection{The Robust Anomaly Detection (RAD) algorithm}

The first phase of devising the anomaly detection algorithm is to extract the stable structure of the cyber-physical system. Let $\mathbf{x}_i$ denote a $d$ dimensional system state vector, representing the reading of the all the sensor and actuators at $i^{th}$ time stamp. The data matrix $M$ is formed by stacking $\mathbf{x}_i$s as rows for $m$ time instances. Therefore, $M$ is a multivariate time series data where each column represents time series reading of a particular sensor or actuator and each row represents the reading of all the sensors and actuators at a particular time stamp. The data matrix $M$ is $(m*d)$ dimension where $d$ is the total number of sensors and actuators and $m$ is the length of time series. 

\begin{equation*}
M=
\begin{bmatrix}
x_{11} & x_{21} & x_{31} \dots & x_{d1}\\
x_{12} & x_{22} & x_{32} \dots & x_{d2}\\
x_{13} & x_{23} & x_{33} \dots & x_{d3}\\
\dots\\
\dots\\
\dots\\
x_{1m} & x_{2m} & x_{3m} \dots & x_{dm}
\end{bmatrix}
\end{equation*}

Each matrix element $x_{ij}$ represents the reading of the $i^{th}$ sensor at $j^{th}$ time step.

\subsubsection{Approximating stable behaviour of the system using robust principal component analysis}

Since the cyber-physical system has a stable behavior under normal operating circumstance, the data observed by the system state vector should lie in a low dimensional subspace. PCA is a widely used technique for dimensionality reduction, but it is highly outlier sensitive. The low dimensional subspace of the data can be determined by Robust PCA\cite{c19}, which will make the intrusion detection mechanism resistant to outliers and poisoning of the training data. To recover the principal components of a matrix $M$ inspite the corruption present in the matrix, we wanted to solve:

\begin{align*}
   \text{minimize} \quad &rank(L)\\
\text{subject to}\quad & M=L+S_0 
\end{align*}

where $S_0$ is a sparse corruption matrix and $L$ is the low rank matrix. But the above problem is non-convex and ill-posed. We instead solve a convex relaxation of the problem, which is called Principal Component Pursuit(PCP). PCP minimizes a weighted combination of the nuclear norm and $l_1$ norm of $S_0$. Solving the relaxed version ensures that we will be able to recover the principal components of the data matrix even if some of the entries in the matrix are corrupt or outliers. 

\subsubsection*{PCP optimization problem}

PCP solves the following convex optimization problem in order to perform RPCA:
\begin{align*}
   \text{minimize} \quad &||L||_* + \lambda ||S_0||_1 \\
\text{subject to}\quad &L+S_0=M 
\end{align*}

The above optimization problem is a convex relaxation of finding low-rank $L$ and sparse $S_0$ because:
\begin{itemize}
    \item The nuclear norm of the $L$ matrix is the sum of the singular values of $L$. The number of non-zero singular values is the rank of a matrix. By minimizing the nuclear norm, we try to indirectly ensure $L$ is low rank.
    \item Minimizing the $l_1$ norm of the matrix $S_0$, indirectly ensures sparsity of $S_0$.
\end{itemize}

The convex PCP problem is solved using augmented Lagrangian multiplier method.

\subsubsection{Representation of stable behaviour of the Cyber-Physical System}

 We obtain the low rank matrix $L$, let us denote each row the matrix by $L_i$, which is the low rank representation of the data point $\mathbf{x_i}$. The basis vectors of $L$ span a low dimensional subspace say $S$. The basis vectors are ortho-normalized using Gram–Schmidt orthogonalization method. Assume, $\{\mathbf{v}_1,\mathbf{v}_2,\cdots,\mathbf{v}_k\}$ be the set of orthonormal basis vectors derived by applying Gram-Schmidt method. This orthonormal basis vectors are organized in the form of column vectors into a matrix $A$.\\
 
 The orthogonal projection of a vector $\mathbf{x}$ on the subspace $S$, spanned by orthonormal vectors of the matrix $A$ is defined by:
%$\mathbf{x} \in {\rm I\!R}^{n}$ can be expressed as $\mathbf{x}=\mathbf{v}+\mathbf{w}$ where $\mathbf{v}$ is and $\mathbf{w}$ is the orthogonal complement.
\begin{align*}
Proj_{A}(\mathbf{x})=A(A^TA)^{-1}A^T\mathbf{x}=AA^T\mathbf{x}
\end{align*}
In the projection operator we have used the orthonormal property of the basis vectors $\{\mathbf{v}_1,\mathbf{v}_2,\cdots,\mathbf{v}_k\}$, giving $A^TA=I_k$. Using the projection operator $Proj_{A}(\cdot)$, the training vectors $\{\mathbf{x}_i\}_{i=1}^m$ are projected into low-dimensional subspace $S$, where vector $\mathbf{x}_i$ consists of reading actuators and sensors at $i^{th}$ timestamp.
For each $\mathbf{x}_i$ in the training dataset, let \begin{align*}
    \mathbf{z}_i=Proj_A(\mathbf{x}_i)=A(A^T\mathbf{x}_i)
\end{align*}
where $\mathbf{z}_i$ is the projection of $\mathbf{x}_i$ in the subspace $S$. The $\mathbf{x}_i$'s are projected in the low-dimensional subspace because the $Proj_{A}(\mathbf{x}_i)$ is the closest point of $\mathbf{x}_i$ in the subspace $S$. 

We then take the geometric median of the projected points:
\begin{align*}
    \mathbf{m}= \arg \min_{\mathbf{y}} \sum\limits_{i=1}^{m} ||\mathbf{z}_i -\mathbf{y}||_2
\end{align*}
Weiszfeld algorithm is a standard tool to compute the above median. The stable behavior of the cyber-physical system is represented by the median $\mathbf{m}$. In order to detect the state of the system in the run-time, we need to project a state vector into the low-dimensional space S, then compute its euclidean distance from the pre-computed median $\mathbf{m}$.\\

\subsubsection{Determining the value of threshold $\theta$}
The geometric median, $\mathbf{m}$ of the projected data points can be considered as a robust central tendency measure for the projected training data points. We set the value of the threshold as highest distance of low rank representation of any point in the training data $L_i$ from the geometric median
\begin{align*}
 \theta &=\max_{i \in [m]} ||\mathbf{m}-L_i||_2
\end{align*}
 In the low dimensional subspace S, the hyper-sphere centered at median $\mathbf{m}$ with radius equals threshold $\theta$, is the region of stable behaviour for the cyber-physical system. If projection of any data point at the test-time lies outside this hyper-sphere, it is flagged as an anomaly. 
\subsubsection{Classification of the test data point}
The stable behavior of the cyber-physical system is parameterized by the triplet $\{A,\mathbf{m},\theta\}$. Consider a test point $\mathbf{x_j}$, RAD computes the anomaly score $a_j$ using:
\begin{align*}
 a_j &=||\mathbf{m}-A(A^T\mathbf{x}_j)||_2\\
\end{align*}
and compares it with the threshold $ \theta $. If the anomaly score $a_j$ is higher than the threshold $\theta$, then the data point $\mathbf{x_j}$ is flagged as an anomaly.

\subsection{Time \& Space Analysis}

We consider the time and space requirement for deploying the proposed RAD intrusion detection system. Since we need to store the triplet $\{A,\mathbf{m},\theta\}$, we shall require space $O(d \cdot r)$, $A$ being a $d\times r$ dimensional and $\mathbf{m}$ being $d$ dimensional. Apart from that, we shall require to store an $r$ dimensional intermediate projection vector $A^T\mathbf{x_j}$ and a $d$ dimensional test data point $\mathbf{x_j}$. Hence the overall space complexity of the deployment is $O(d \cdot r)$, where $r$ is the rank of the low dimensional subspace $S$. Consider the steps involved in classifying a test data point:

%\begin{itemize}
    %\item Matrix $A$ of dimension $d\times r$ , where $d$ is the number features in the state vector $\mathbf{x_i}$ and 
    %\item An intermediate result $A^T\mathbf{x}_i$ of dimension $r$.
    %\item The median vector of dimension $d$ and a scaler $\theta$.
%\end{itemize}

\begin{itemize}
    \item To compute the projection $A(A^T \mathbf{x}_j)$of a data point $\mathbf{x}_j$, we need two matrix vector multiplication. Thus, time complexity of the projection step is $O(d \cdot r)$. 
    \item Computing the anomaly score would require $O(d)$ time, as median $\mathbf{m}$ is $d$ dimensional.
\end{itemize}

With comparison with threshold taking $O(1)$ time, the time complexity of our proposed IDS is $O(d \cdot r)$. Since, $r$ can be considered constant, the proposed method is said to work in $O(d)$ space and time complexity. 

\section{Experiments \& Results}

To evaluate the performance and accuracy of the proposed intrusion-detection model, we have used the SWaT dataset, Tennessee Eastman process dataset and a dataset corresponding to a Government sponsored power distribution testbed. Note that the training data contains both normal and anomalous data points, so that algorithms that assume training on normal condition are at a disadvantage. 

\subsection{Training Phase}
\subsubsection{Secure Water Treatement(SWaT) Testbed}

The SWaT dataset\cite{c9} comprises of reading  from the sensors and actuators of an operational water treatment testbed, at different timestamp. The testbed comprises of 6 different stages, where each stage signifies a water treatment process. The physical parameters that are measured by the sensors are level of the water in the tanks, differential pressure, flow indicators etc. The actuators are different kind of valves that can either be ON or OFF. The programmable logic controllers (PLC) take the sensor reading as input and give commands to the actuators accordingly. If the attacker can manipulate the sensor readings then PLC will give commands that are undesirable, resulting into an anomalous behavior of the cyber-physical system. Commands can be also given through SCADA and the overall state of the system is maintained at SCADA. Manipulating the sensors and actuators data in the SCADA will also have an undesired effect on the system.

The SWaT dataset $\{\mathbf{x}_i,y_i\}_{i=1}^{m}$, $\mathbf{x}_i \in {\rm I\!R}^{d}$, has $m=449919$ and $d=51$. The vector $\mathbf{x}_i$ consists of reading of actuators and sensors at $i$th timestamp. It is a labeled dataset where $y_i$ is the label representing the state of the system either under normal or under attack. Since RAD is designed to work even with unlabelled data, it does not require the labels. However, as a meta information the label tells that $88 \%$ data points are normal and rest $12 \%$ are attack. There are $36$ attack scenarios in this dataset.

We give the data matrix $M$ which has $\mathbf{x_i}$'s as rows, as an input to the PCP algorithm to recover the low rank matrix $L$. The rank $r$ of $L$ is found to be $18$. Figure \ref{fig:rpca_error} shows the error $||M-L-S_0||_F$ where is the Frobenius norm of the matrix $M-L-S_0$ is considered, as the constraint in the optimization problem was $L+S_0=M$. The stopping criteria is set to an error less than $10^{-7}||M||_F$. PCP required $5970$ epochs to perform the RPCA.

\subsubsection{Tennessee-Eastman Process}

The Tennessee-Eastman process data is the dataset generated for the research purpose of anomaly detection and fault diagnosis of industrial control systems by simulating a chemical plant\cite{c21}. The normal operation of the chemical plant is simulated as well as attacks are simulated in the dataset by manipulating a targeted set of variables.

In the TE process dataset $\{\mathbf{x}_i\}_{i=1}^{4801}$ , $\mathbf{x}_i \in {\rm I\!R}^{41}$, the vector $\mathbf{x}_i$ consists readings of sensors at each timestamp. In the TE process dataset $83\%$ data points belongs to normal and the rest are observations corresponding to anomalous behavior of the system. The data matrix $M$ is of dimension $(4801*41)$, was given to PCP to find a low rank matrix $L$ of rank $r=24$. Figure \ref{fig:rpca_error} shows the corresponding error at each epoch.
The matrix $A$ comprising of the orthonormal basis vectors of $L$ is of dimension $(41*24)$. 

%The median is a $(41*1)$ column vector.

\subsubsection{Govt. Sponsored Power Testbed }

The power testbed is a three phase power distribution testbed in which several loads are connected like industrial automation testbed, conveyor belt and water treatment testbed. The various physical parameters like voltages, current and power are measured with the help of sensors and remote terminal unit (RTU) and are sent to the SCADA from PLC by using industrial communication protocols like MODBUS where the overall state of the system is maintained.

The power testbed dataset $\{\mathbf{x}_i\}_{i=1}^{12310}$ has $\mathbf{x}_i \in {\rm I\!R}^{31}$. The vector $\mathbf{x}_i$ consists readings of sensors corresponding to voltage, current, power and phase angle readings. The first $12125$ data points of this dataset belongs to normal behavior of the system and  the rest $185$ are observations of the system under attack.

\begin{figure}[ht!]
\centering
\includegraphics[width=\linewidth]{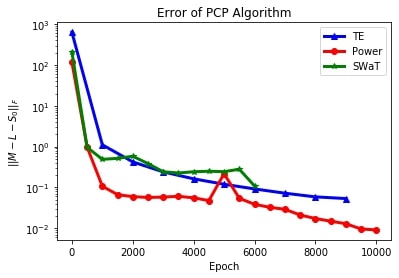}
\caption{Error $||M-L-S_0||_F$ calculated after each epochs for Principal Component Persuit algorithm on SWaT, Tennessee and Power dataset. The decreasing error shows that algorithm successfully recovers a low rank matrix $L$}
\label{fig:rpca_error}
\end{figure}

The data matrix $M$ is of dimension $(12310*31)$ is given as an input to the PCP algorithm to recover the low rank matrix $L$. The rank $r$ of $L$ is found to be $9$. Figure \ref{fig:rpca_error} shows the corresponding error after each epoch. The matrix $A$ comprising of the orthonormal basis vectors of $L$ is of dimension $(31*9)$. The median is a $(31*1)$ vector.

\subsection{Test Phase}
For demonstration, we show how proposed intrusion dection system RAD detects the attack under different attack scenarios, for three datasets. We compare the performance with other comparable methods.

 \subsubsection{Secure Water Treatement(SWaT) Testbed}

%After the training, we shall store only three parameters, namely $A, \mathbf{m}, \theta$ for the functioning of the intrusion detection model in test phase. This shows that the intrusion detection model designed by our algorithm does not require much memory and can be stored in a resource constrained environment. 

\subsubsection*{Attack Scenario 1}

\begin{figure}[ht!]
\centering
\includegraphics[scale=0.45]{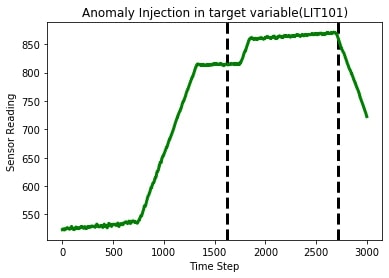}
\caption{Sensor LIT-101 showing ongoing attack}
\label{fig:read1}
\end{figure}

In attack scenario 1, the attacker's intent is to overflow the tank 1 by attacking on the actuator MV-101 to remain open even if the level indicator sensor LIT-101 is above the highest permitted level. Figure \ref{fig:read1} shows the reading of the sensor LIT-101 at each time step. The highest permitted level is $800mm$. We can see that the attacker is successfully overflowing the tank from the time step $1600$ to $2600$. This is denoted by the vertical bars which signifies starting and stopping timestamp of the attack. After the attack the level of tank comes within the permitted level.

\begin{figure}[ht!]
\centering
\includegraphics[scale=0.45]{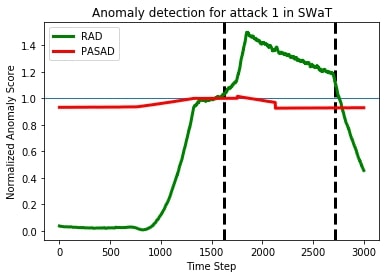}
\caption{Normalized anomaly score at each time step shows RAD successfully identifying the attack, while PASAD fails to identify it.}
\label{fig:attack1}
\end{figure}

Figure \ref{fig:attack1} shows the corresponding anomaly score at each time step. The blue horizontal line shows the normalized threshold. It is normalized to one, since anomaly scores are divided by actual threshold of that system, to make different anomaly scores comparable. It can be observed that RAD correctly identifies the start of the attack, as the anomaly score exceeds the threshold. Again, when the attack stops and the level of the water in the tank comes within the permitted level the anomaly score drops below threshold. This depicts RAD successfully identifying the system behavior, however PASAD fails to flag the anomaly because it's anomaly score stays below the threshold.

\subsubsection*{Attack Scenario 2}

\begin{figure}[ht!]
\centering
\includegraphics[scale=0.45]{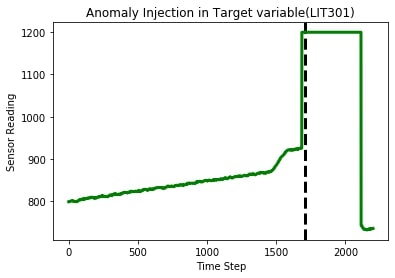}
\caption{Sensor LIT-301 reading at each time step}
\label{fig:read2}
\end{figure}

In attack scenario 2, the attacker's intent is to overflow the tank 3 by attacking on the sensor LIT-301 by increasing the level of the tank suddenly well above the maximum level. The attacker did not stop and the tank does overflow($>900mm$) for a period of time. \ref{fig:read2} shows the reading of the sensor LIT-301 at each time step. The highest permitted level is 900. We can see that the attacker is successfully overflowing the tank. After the attack the level of tank comes within the permitted level.

\begin{figure}[ht!]
\centering
\includegraphics[scale=0.45]{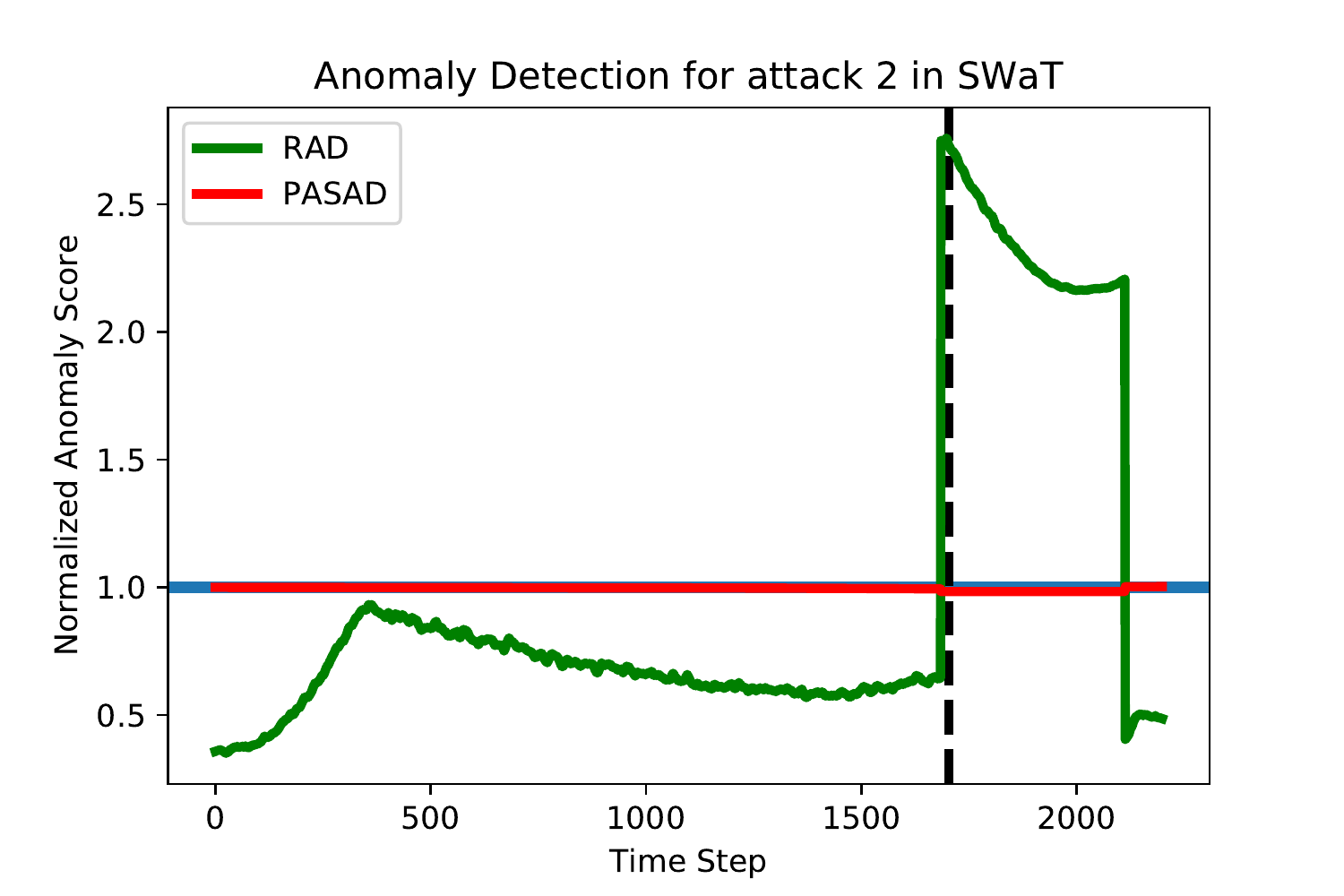}
\caption{Sharp rise in RAD anomaly score capturing the attack.}
\label{fig:attack2}
\end{figure}

Figure \ref{fig:attack2} shows the anomaly score at each time step with the blue horizontal line as the threshold. RAD anomaly score captures the attack started and eventually when the level of water is above the permitted level, it crosses the threshold. After the attack, when level of the water comes within the permitted level, the RAD anomaly score drops below the threshold. The sharp drop in the anomaly score corresponding to the sharp drop in the sensor reading describes that the intrusion detection model can capture the deviation in the system behavior. On the contrary, PASAD anomaly score does not reflect the attack.

 \subsubsection{Tennessee Eastman Process Dataset}
 
 %After the training phase there are three learning parameters that are stored i.e, $A$ of $(41*24)$ dimension, median vector of $(41*1)$ dimension and a scalar threshold value $\theta$ for the functioning of the intrusion detection model in test phase. 
 
 \subsubsection*{Attack Scenario 1}

\begin{figure}[ht!]
\centering
\includegraphics[scale=0.45]{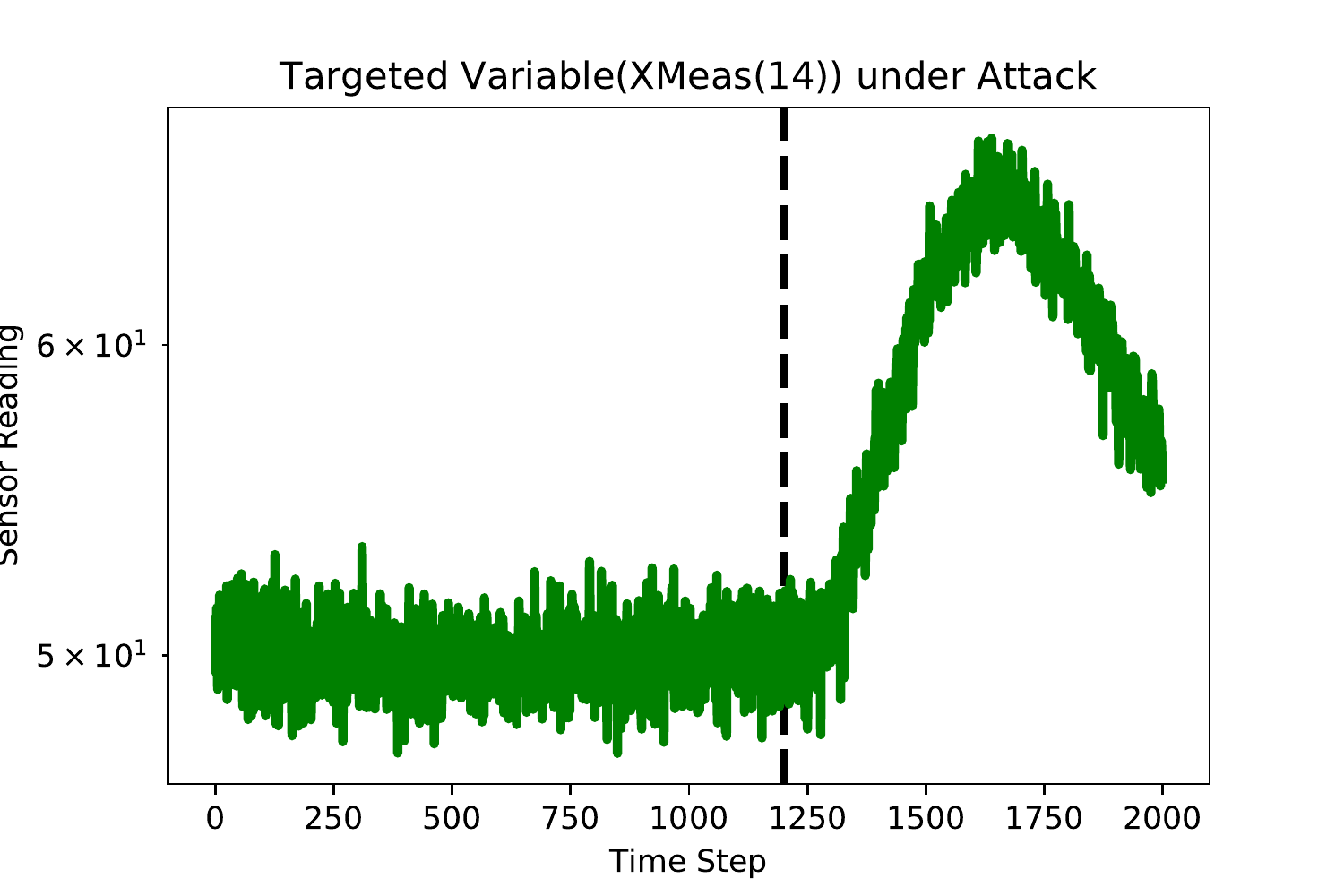}
\caption{Xmeas(14) Sensor reading at each Time Step}
\label{fig:read1_te}
\end{figure}

 In attack scenario 1, the attacker's intent is to increase the value of the sensor abruptly by gradually increasing to a point well above the desired level. Figure \ref{fig:read1_te} shows the reading of the sensor XMeas(14) with time when the sensor is under attack.

\begin{figure}[ht!]
\centering
\includegraphics[scale=0.45]{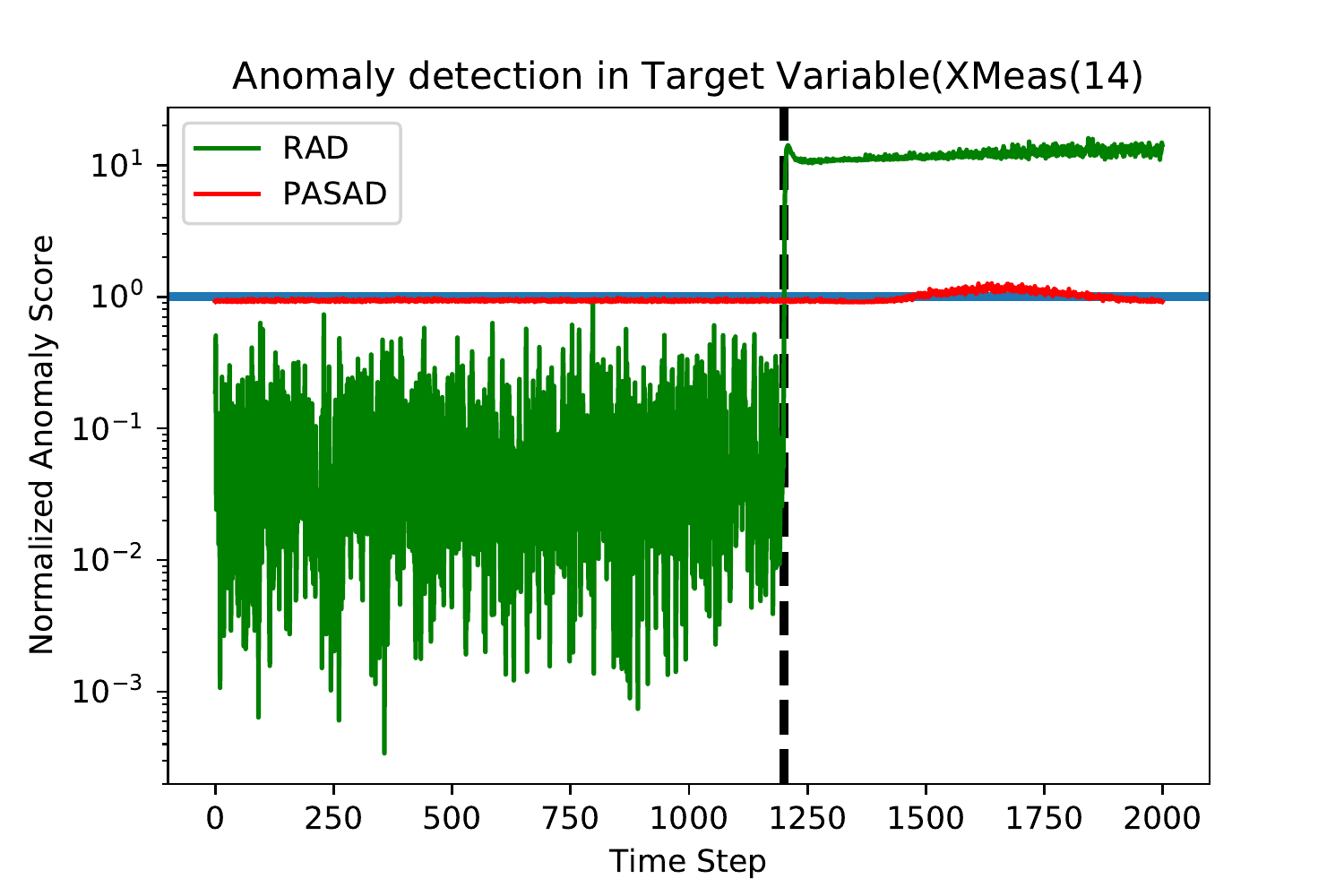}
\caption{Anomaly Score at each Time Step}
\label{fig:attack1_te}
\end{figure}

\subsubsection*{Attack Scenario 2}

\begin{figure}[ht!]
\centering
\includegraphics[scale=0.45]{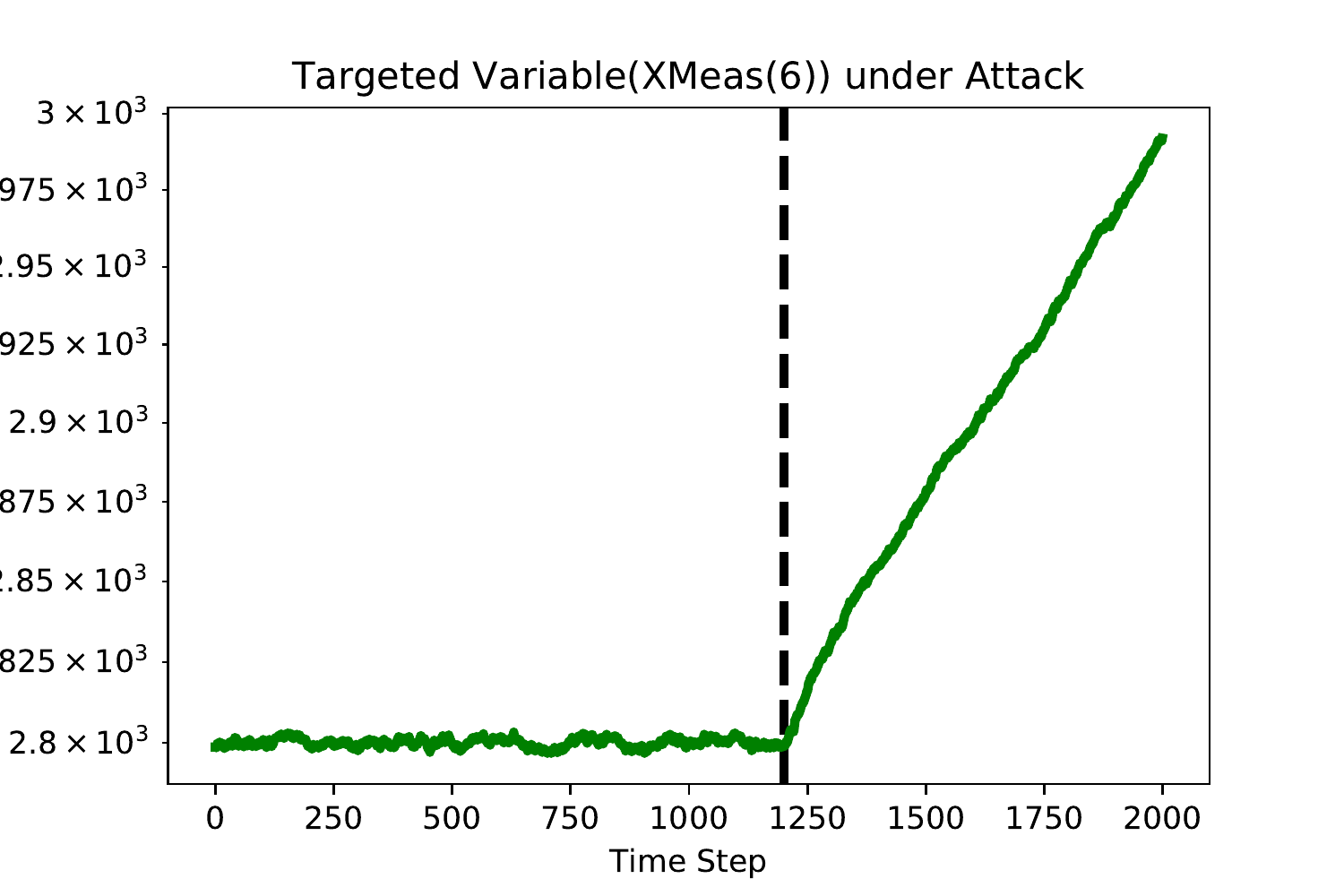}
\caption{Xmeas(6) Sensor reading at each Time Step}
\label{fig:read2_te}
\end{figure}

 In attack scenario 2, the attacker's intent is to increase the value of the sensor abruptly by gradually increasing to a point well above the desired level. Figure \ref{fig:read2_te} shows the reading of the sensor XMeas(6) with time when the sensor is under attack.

\begin{figure}[ht!]
\centering
\includegraphics[scale=0.45]{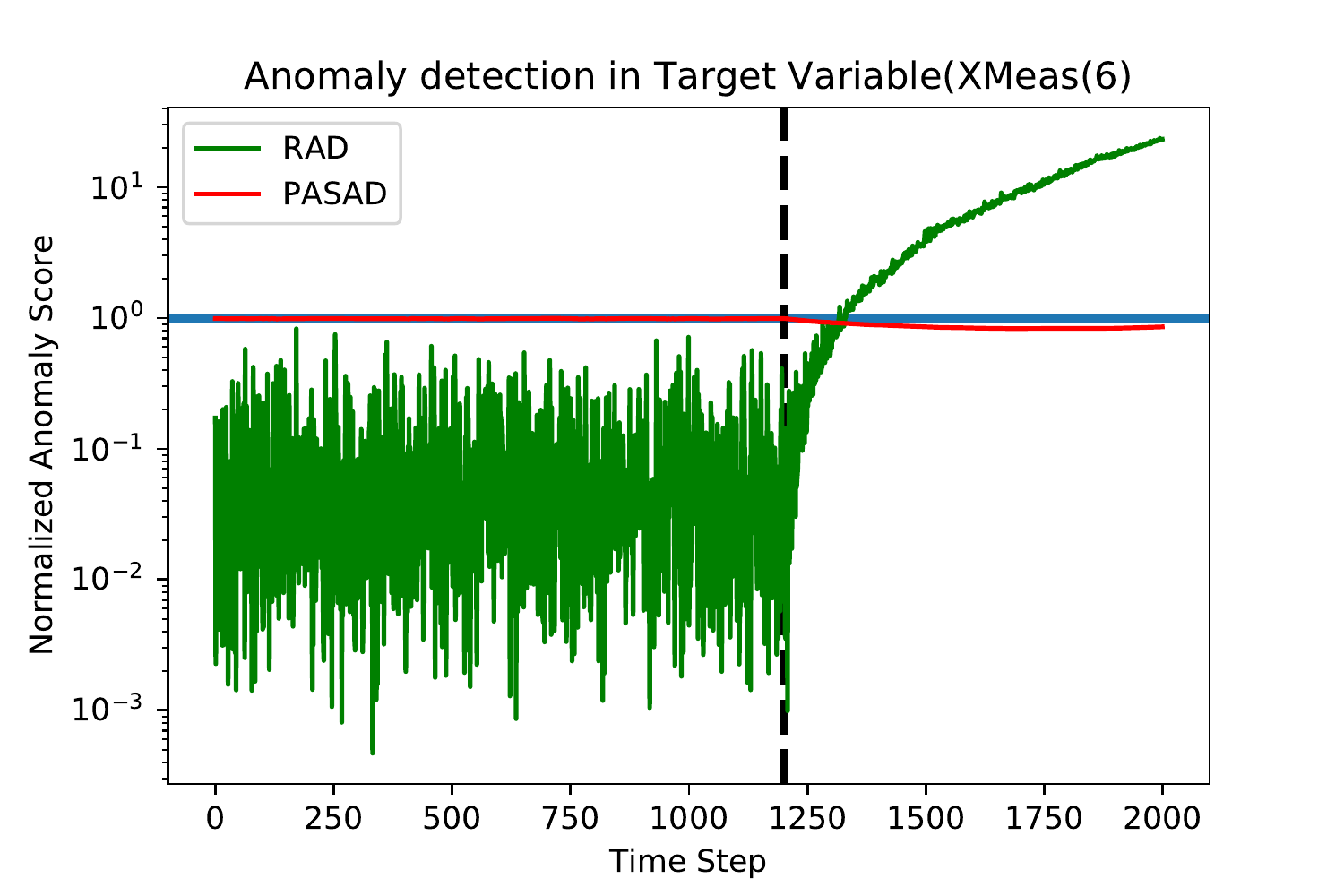}
\caption{Anomaly Score at each time step}
\label{fig:attack2_te}
\end{figure}

Figure \ref{fig:attack1_te} and Figure \ref{fig:attack2_te} show the anomaly score at each time step, with the blue horizontal line as the threshold. It clearly shows that when the attack starts and eventually when the sensor reading is above the permitted level the RAD anomaly score crosses the threshold flagging the anomaly of the cyber-physical system. There is delayed detection by PASAD in the attack scenario 1, but it for fails to detect the attack in scenario 2.

\subsubsection{Power Distribution Dataset}

 %After the training phase there are three learning parameters that are stored i.e, $A$ of $(31*9)$ dimension, median vector of $(31*1)$ dimension and a scalar value $\Theta$ for the functioning of the intrusion detection model in test phase. 

\begin{figure}[ht!]
\centering
\includegraphics[scale=0.45]{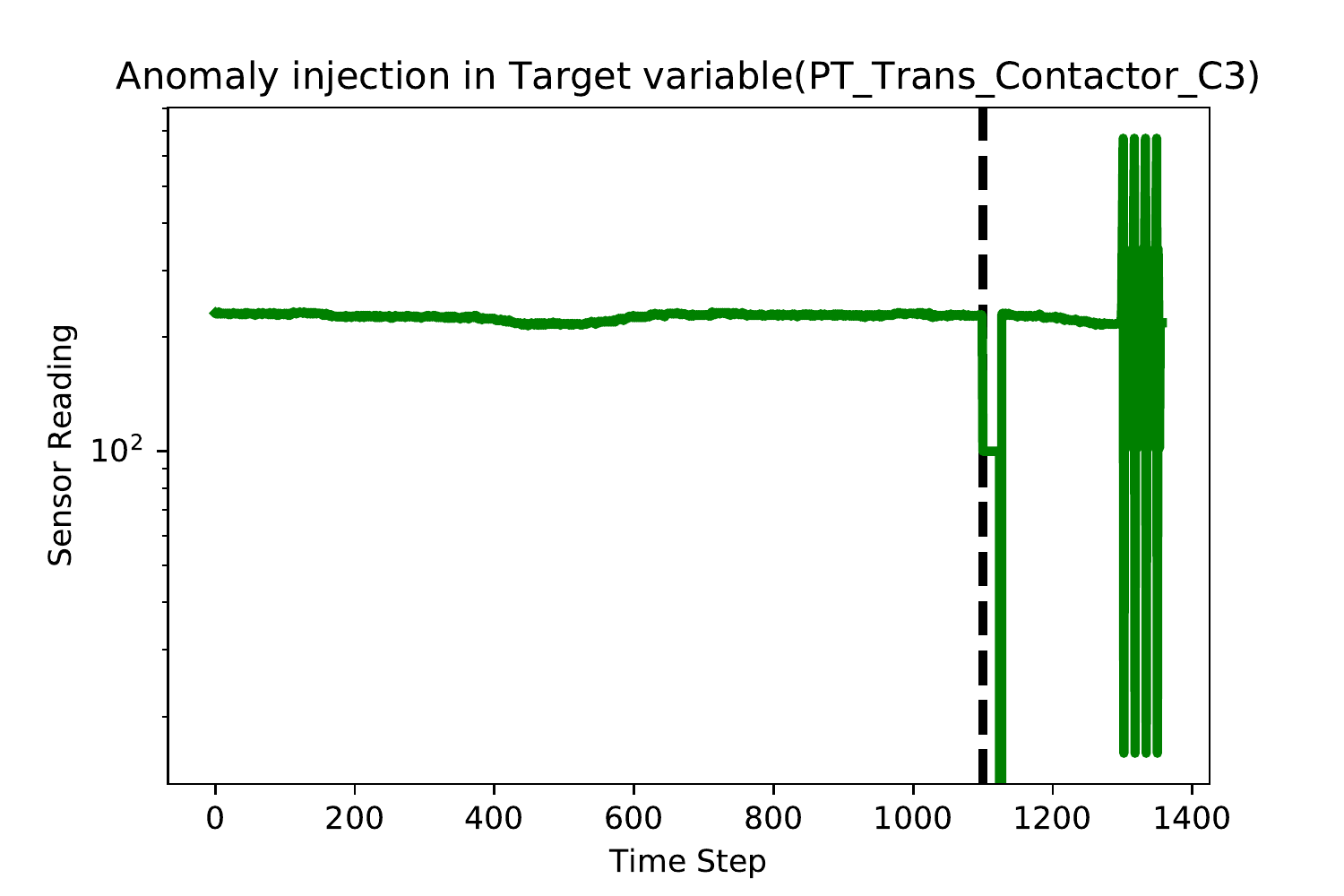}
\caption{ PT transcontractor C3 sensor reading under attack}
\label{fig:read2_pd}
\end{figure}

In this attack scenario, the attacker's intent is to increase the value of the sensor `PT trans-contractor C3', which is measures trans-conductance in the third phase of a 3-phase power distribution test bed. The attacker suddenly generates a spike in the reading. Figure \ref{fig:read2_pd} shows the reading of the sensor XMeas(6) when the sensor is under attack.

\begin{figure}[ht!]
\centering
\includegraphics[scale=0.45]{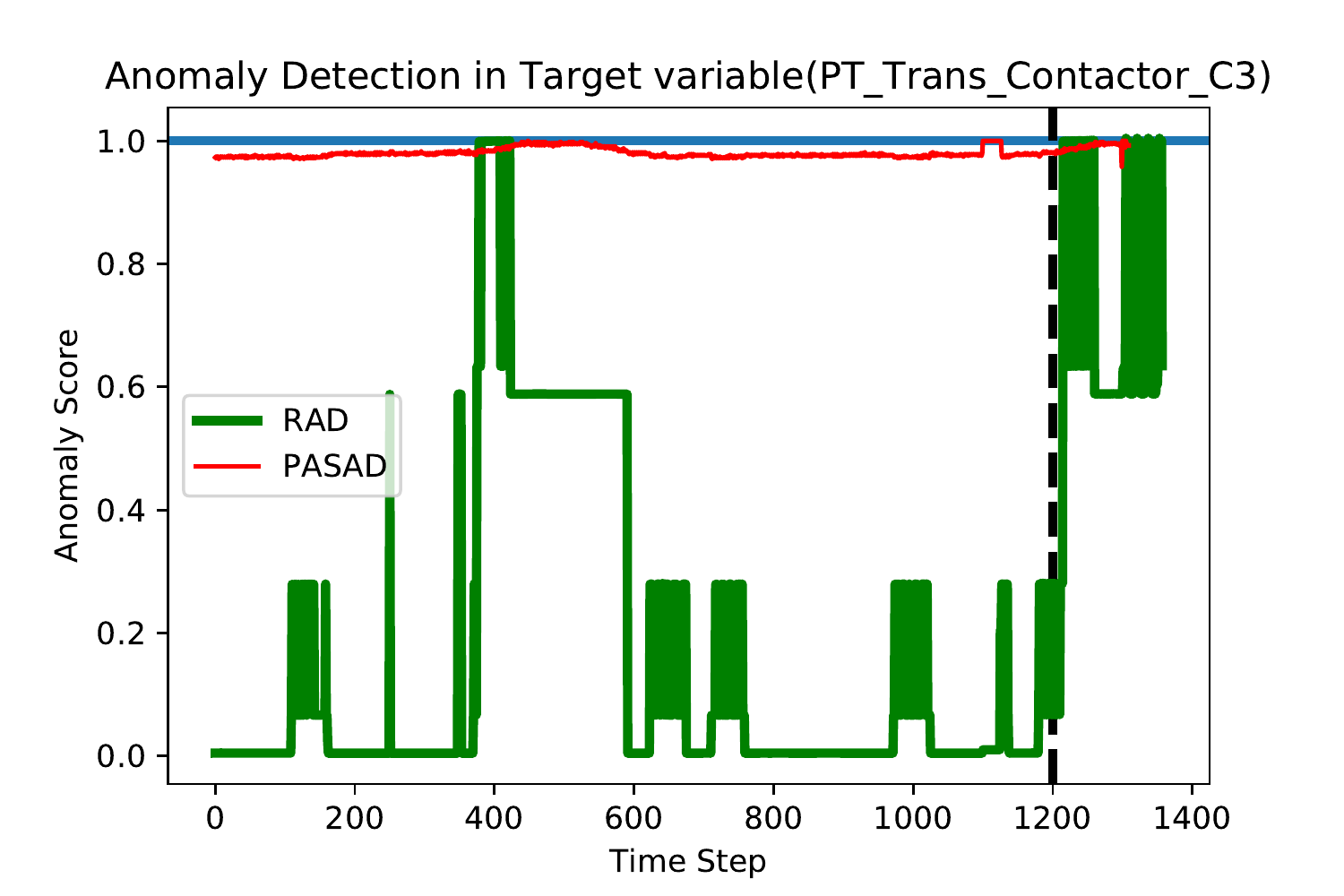}
\caption{Anomaly Score at each Time Step}
\label{fig:attack2_pd}
\end{figure}

Figure \ref{fig:attack2_pd} shows the anomaly score at each time step with the blue horizontal line as the threshold. It shows that with a sudden abnormality in the behavior of the cyber-physical system, the RAD anomaly score spikes, and touches the threshold. PASAD anomaly score does not respond to the attack.

\subsection{Robustness in the presence of outliers in the training data}

In order to demonstrate robustness of RAD, we perform experiment where we introduce two kinds of corruption additionally into the training data which already contains the normal and attack scenario as described before.
\begin{itemize}
    \item Gaussian Noise: We introduce $N(0,1)$ Gaussian noise to the whole training data.
    \item Burst Outliers: Few variables are corrupted at regular interval consecutively for some time interval.
\end{itemize}

\subsubsection*{Tennessee Eastman Process Dataset}

The variables XMeas(6) and XMeas(14) are corrupted at a regular interval of $25$ time steps. Additionally Gaussian noise was introduced to the entire training data.

\begin{figure}[ht!]
\centering
\includegraphics[scale=0.45]{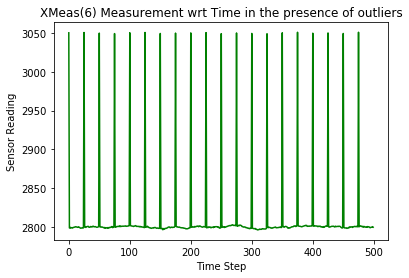}
\caption{Xmeas(6) sensor reading after outlier injection in training}
\label{fig:read_o_te}
\end{figure}

Figure \ref{fig:read_o_te} shows the reading of a variable Xmeas(6) after outliers are injected at regular intervals. The spikes in the reading depicts the outliers in the varaible. Similar ouliers were introduced in the XMeas(14).

\begin{figure}[ht!]
\centering
\includegraphics[scale=0.45]{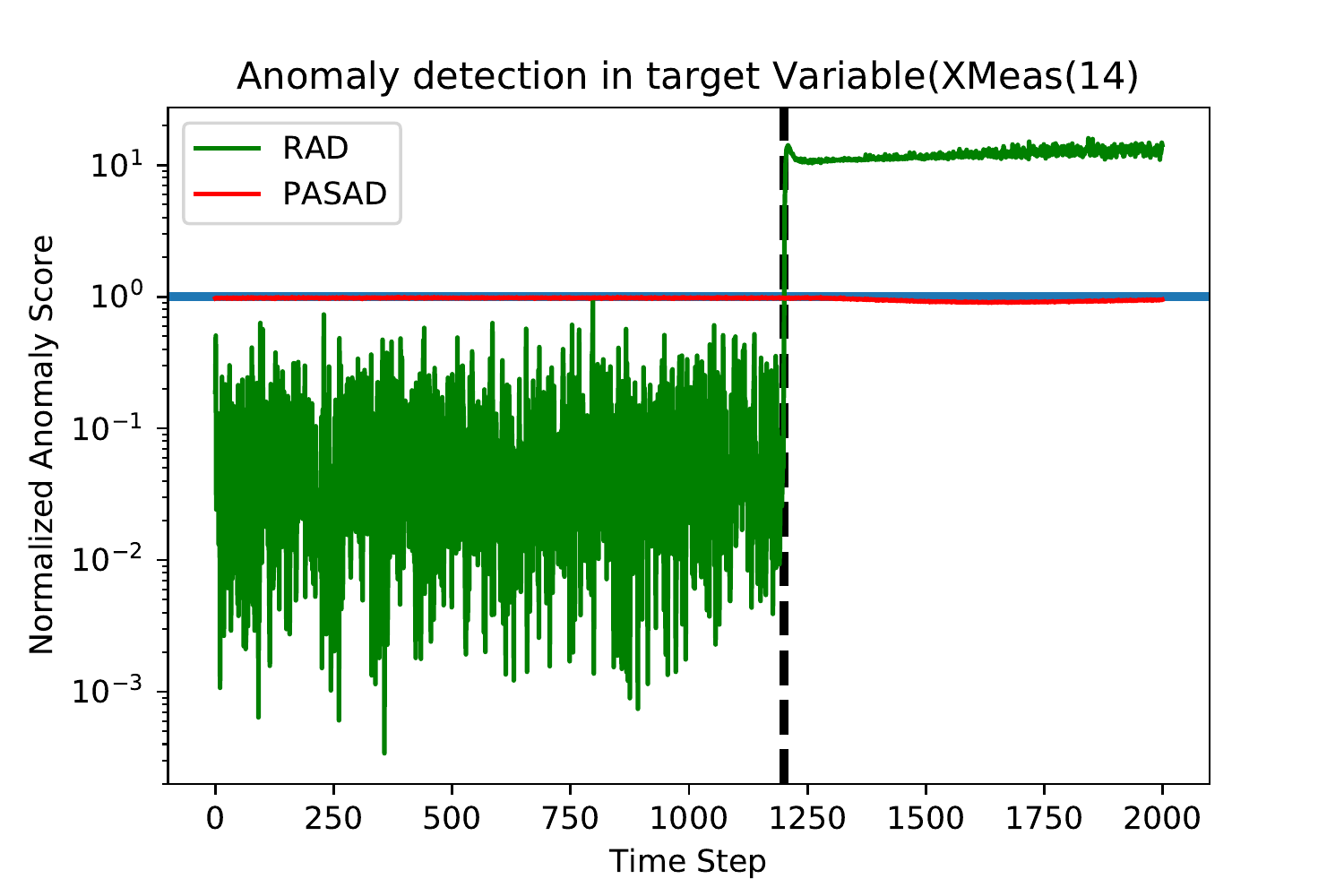}
\caption{Anomaly score corresponding to attack in the test time on variable XMeas(14)}
\label{fig:attack_o_te}
\end{figure}

At the test time the variable XMeas(14) was attacked as depicted earlier in the Figure \ref{fig:read1_te}. Figure \ref{fig:attack_o_te} shows the RAD anomaly score at the corresponding time steps. Successful detection of attack shows the robustness of RAD to detect the anomalies, even though additional outliers in the form of Gaussian noise or random jitter, were present in the training data. Comparing to \ref{fig:attack2_te} we can see the PASAD anomaly score have worsened, due to further training data corruption and it is below the threshold.

%The rank $r$ of the recovered low rank matrix $L$ is $24$ with out the injection of outliers in the training data, and rank $r$ of the recovered low rank matrix $L$ is $23$ with the injection of outliers in the training data.

\section{Real-time Deployment and Memory Requirement}
As discussed in Section 4.2, in order to calculate anomaly score to generate alarms in the test time,
we shall require parameters $\{A,\mathbf{m},\theta\}$ and vectors $\{A^T \mathbf{x}_i, \mathbf{x_i}\}$. Considering the power distribution dataset, where $d=31$ and $r=9$. The matrix $A$ be $(31*9)$ dimensional and scalar is $4$ bytes, the memory required to store the parameters, is $(31*9+31*2+9+1)*4=1404$ bytes, which is $1.404$KB. 
The general-purpose machines in which the SCADA software operates have a RAM of $256$MB and running simple code for matrix multiplication, and a conditional operation is feasible. Therefore, the proposed IDS is feasible to be deployed in real infrastructure within the SCADA host to detect attacks in real-time.

Generally, Programmable Logic Controllers(PLC) have Internal RAM of $4$MB. To store a single floating point value in the BMXRMS008MP Schnieder PLC requires a 16bit word i.e, is 2bytes. The memory required to store our IDS in this particular PLC is $702$ bytes, which is less than a KB. If matrix multiplication and comparison of two variables are allowed in the PLC, which is generally possible using STL language, then our proposed IDS can be deployed within PLC.

\section{Conclusion}

Developing intrusion detection system for cyber-physical systems is necessary because of their critical nature and vulnerability to attacks. Slight change in their behavioral dynamics can significantly impact the real-world installations. Moreover, a live cyber-physical system can not be trusted to be operating under desired normal condition. Therefore, a robust training method for IDS is necessary. The proposed Robust Anomaly Detection (RAD) method tolerates corruption and attacks in the training data and successfully detects an ongoing attack in the cyber-physical system in $O(d)$ time. This paper experimentally demonstrates the performance of the proposed IDS in multiple data set by considering different attack scenario. It can be deployed efficiently within the SCADA host as well as PLC using matrix multiplication.


\ifCLASSOPTIONcompsoc
\else
 % \section*{Acknowledgment}
\fi




%\bibliographystyle{IEEEtran}

\end{document}